\documentclass[12pt]{iopart}
\usepackage{epsf}
\usepackage{graphicx}

\newcommand{\p}{{\rm p}}
\newcommand{\s}{{\rm s}}
\newcommand{\gcc}{{\rm \,g\,cm^{-3}}}
\newcommand{\kms}{{\rm \,km/s}}
\newcommand{\K}{{\rm \,K}}
\renewcommand{\etal}{{\em et$\,$al. }}

\begin{document}
{\small \vspace*{-7mm} \noindent Submitted to proceedings of the conference on Strongly Coupled Coulomb Systems 2002 in Santa Fe, published in {\em J. Phys. A} with guest editors J. Dufty, J.F. Benage, and M.S. Murillo.\vspace*{-5mm}}

\title[B. Militzer: Calculation of shock Hugoniot curves of precompressed liquid deuterium]
       {Calculation of Shock Hugoniot Curves of Precompressed Liquid Deuterium}

\author{Burkhard Militzer\dag\
\footnote[3]{militzer@llnl.gov}
}

\address{\dag\ Physics Department,
              Lawrence Livermore National Laboratory,
              University of California, Livermore, California 94550, USA}

\begin{abstract}
Path integral Monte Carlo simulations have been used to study
deuterium at high pressure and temperature. The equation of state has
been derived in the temperature and density region of $10\,000 \leq T
\leq 1\,000\,000\,$K and $0.6 \leq \rho \leq 2.5\,$g$\,$cm$^{-3}$. A
series of shock Hugoniot curves is computed for different initial
compressions in order to compare with current and future shock wave
experiments using liquid deuterium samples precompressed in diamond
anvil cells.
\end{abstract}






The equation of state of dense deuterium has been the topic of intense
discussions since Nova laser shock wave experiments~\cite{Si97}
measured the Hugoniot curve up to megabar pressures in 1997 and
predicted an unexpectedly high compressibility. Four years later,
Knudson \etal measured a significantly smaller compressibility using
magnetically driven flyer plates~\cite{Kn01}. This controversy
continues to initiate new efforts to determine the EOS in different
regions of the high temperature phase diagram. In a new approach to
reach higher densities, G.W.~Collins and P.~Cellier use a diamond anvil
cell (DAC) to precompress samples of liquid deuterium and then launch
laser shock into them. The focus of this article is to use path
integral Monte Carlo (PIMC) results to determine Hugoniot curves for
precompressed samples in order to make a prediction for comparison
with current and future experiments.

The two series of shock wave experiments using the Nova
laser~\cite{Si97,Co98} reached pressures of up to 340$\,$GPa and
predicted a 50\% higher compressibility than previously estimated. The
results provided the first experimental data in a regime of extreme
pressure and temperature, in which our understanding had so far been
primarily based on analytical models and computer simulations. The
experimental findings suggested that standard EOS models such as
Sesame~\cite{Ke83} were too {\em stiff}, predicting a maximum
compression of about 4-fold the initial density.  Instead the measured
EOS was found to be closer to {\em softer} models like~\cite{Ro98}
that predicts a maximum compression ratio of about 6. Such a
difference in the EOS of dense hydrogen and deuterium would have
significant consequences for our understanding of Jovian planets,
brown dwarfs, and low mass stars as well as for the design of the
inertial confinement fusion strategies.

As a consequence of the publication of the Nova measurements, a number
of chemical models were developed which incorporated the experimental
information directly or indirectly to fit unknown model parameters and
then predicted an EOS closer to Nova experiments. This is a common
strategy used to improve such a complex function as the deuterium EOS,
which must describe different phases and regimes and spans many orders
of magnitude in temperature and pressure. Typically the EOS is pieced
together from different analytically known limits, experimental data,
and computer simulation results.
However, the high compressions seen in the predictions of chemical
models were not confirmed by first principles simulation techniques
such as density functional molecular dynamics~\cite{Le97} and
PIMC~\cite{MC00}. Both results are in relatively good agreement with
each other, predicting a lower compressibility and a Hugoniot curve
close to that of the Sesame model.

In a different series of laser shock wave experiments,
Mostovych \etal\cite{Mo00} used a reshock technique to probe the
EOS at higher compressions. The results were also in better agreement
with the soft EOS models and differed significantly from {\em ab
initio} simulations~\cite{Mi01}. It should be noted that the error
bars were larger than those of the Nova results.

The controversial discussion of different results changed dramatically
with the publication by Knudson \etal\cite{Kn01}. Instead of a laser,
the Z pulse power machine was used as magnetic drive to accelerate a
flyer plate launching a shock into deuterium. The achieved pressures
have not yet exceeded 100 GPa but a different trend in the Hugoniot
curve can already be identified. The results clearly predict a
significantly lower compressibility, which is in good agreement with
{\em ab initio} simulations and with other stiff EOS models. Given two
different experimental results, it can be expected that efforts to
resolve this discrepancy will intensify.
So far, no satisfactory explanation has been given despite many
different attempts to interpret the data. This also included
suggestions for the existence of nonequilibrium states with different
temperatures for ions and electrons. Darma-Wardana \etal\cite{DW02}
suggested $T_e<T_i$ while Gygi \etal\cite{GG02} proposed
$T_e>T_i$. The later prediction resulted from a remarkable
computational effort involving the first dynamic simulation of shock
propagation using the Car-Parinello method.

Chemical models have been constructed to predict Hugoniot curves
ranging from 4 and 6-fold compression while {\em ab initio}
simulations, with the exception of wave packet molecular
dynamics~\cite{Kn02}, predict a stiffer Hugoniot. However, it is
generally assumed that the final determination of the EOS will result
from further experimental work. New efforts are on the way and novel
techniques are being developed. While the principle Hugoniot
represents only one line in the phase diagram, other methods can probe
different densities. Double and multiple shock reverberation
measurements can access higher density regimes. Also, a pulse shaping
technique is being tested by Knudson \etal to generate compression
states on an isentrope rather than Hugoniot states. One new way to
access higher densities is to use a DAC to precompress deuterium
before launching a single shock through the diamond into the
sample. This method allows one to choose a different initial state
characterized by $(\rho_0,P_0,E_0)$ leading to shock states
$(\rho_1,P_1,E_1)$ given by~\cite{Ze66},
\begin{eqnarray}
\label{hug_p}
P_1-P_0 &=& \rho_0 \, u_\s \, u_\p~, \\
\frac{\rho_1}{\rho_0} &=& \frac{u_\s}{u_\s-u_\p}~,\\
\label{hug}
0 &=& (E_1-E_0) + \frac{1}{2}\,\left(V_1-V_0\right)\,(P_1+P_0)~.
\end{eqnarray}
For a given EOS, shock velocity $u_\s$ and particle velocity $u_\p$
can be derived from, $\,u_\p^2 = \xi/\eta$ and $\,u_\s^2 = \xi \eta$
with $\xi=(P_1-P_0)/\rho_0$ and $\eta=1-\rho_0/\rho_1$. 

In our calculations, we use $E_0=-15.886\,\rm{eV}$ per atom and $P_0 =
0$ as used in~\cite{MC00}. We expect that there will be small
corrections to the initial state based on the initial pressure and
temperature, which can be applied to our results when the initial
state of a specific experiment is known. Different initial conditions
like $\delta E_0$, caused primarily by a higher initial temperature,
and $\delta P_0$, from the precompression, shift a particular Hugoniot
curve characterized by a fixed $\rho_0$ from density $\rho_1$ to
$\rho_1+\delta \rho_1$. The density correction $\delta \rho_1$ to
first order is given by,
\begin{eqnarray}
\delta \rho_1 &=& \frac{\rho_1^2}{P_1+P_0} \; \frac{2}{m}  \; \delta E_0\;,\\
\delta \rho_1 &=& \frac{\rho_1^2}{P_1+P_0} \left(\rho_1^{-1}-\rho_0^{-1}\right) \; \delta P_0\;.
\end{eqnarray} 
We expect these corrections to be relatively small compared to the
difference between the Nova and Z-pinch results, which correspond
approximately to a difference of 3~eV per atom in the internal energy
or to an EOS change of $-2\,$eV in $PV$ per atom (see~\cite{Mi01}).

The purpose of this article is to make a prediction based on PIMC
simulations for the Hugoniot curves in current and future experiments
using samples of precompressed liquid deuterium. Due to the length
restriction for this article, we cannot describe any details of the
PIMC method, which can be found in~\cite{Ce95,Ce96,Ce91,BM00}. Also
for the EOS efforts using PIMC simulations, we refer the reader to the
literature~\cite{PC94,Ma96,MC01}. For this particular calculation, we
used EOS data published in~\cite{BM00,MC00,Mi01} and results from a
number of additional simulations.

\begin{figure}
\begin{center}
\includegraphics[angle=0,width=13.7cm]{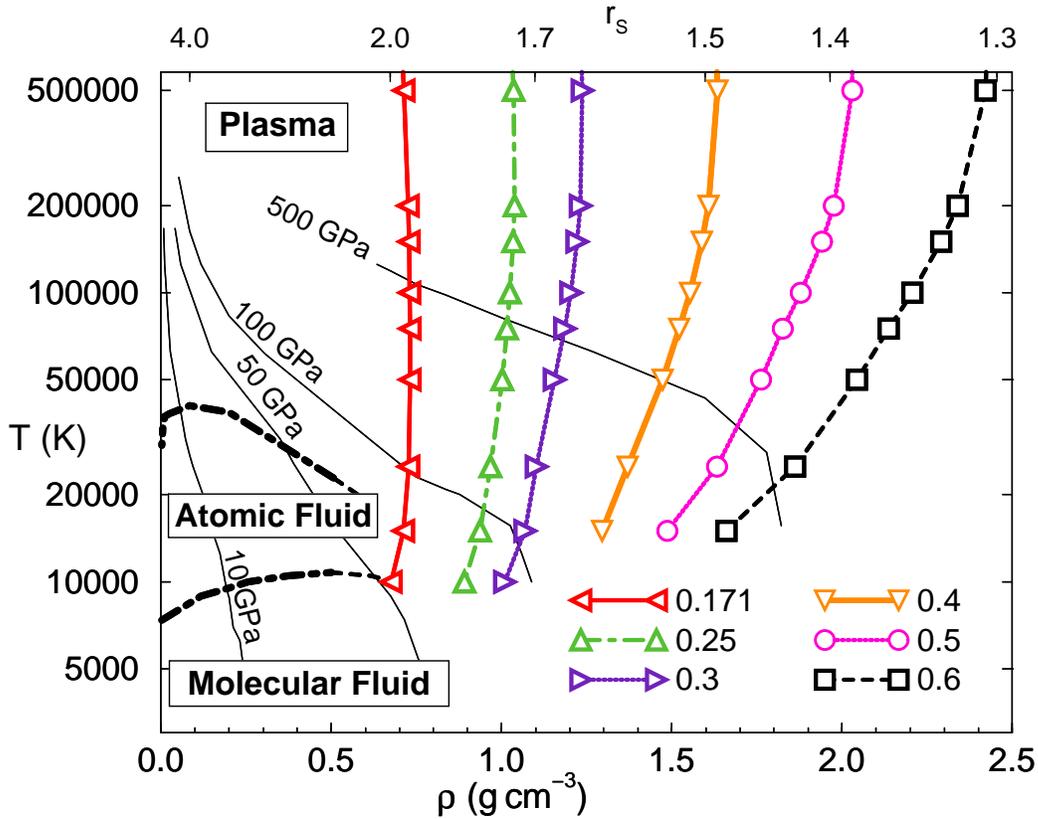}
\end{center}
\caption{\label{phase}
	Deuterium temperature density phase diagram with principle
	Hugoniot curves calculated from PIMC simulations for different
	initial densities $\rho_0$ given in the legend in units of
	$\gcc$. The thin solid lines represent isobars, dash-dotted
	lines indicate the approximate boundaries of the molecular,
	the atomic, and the plasma regime.}
\end{figure}

\begin{figure}
\begin{center}
\includegraphics[angle=0,width=13.7cm]{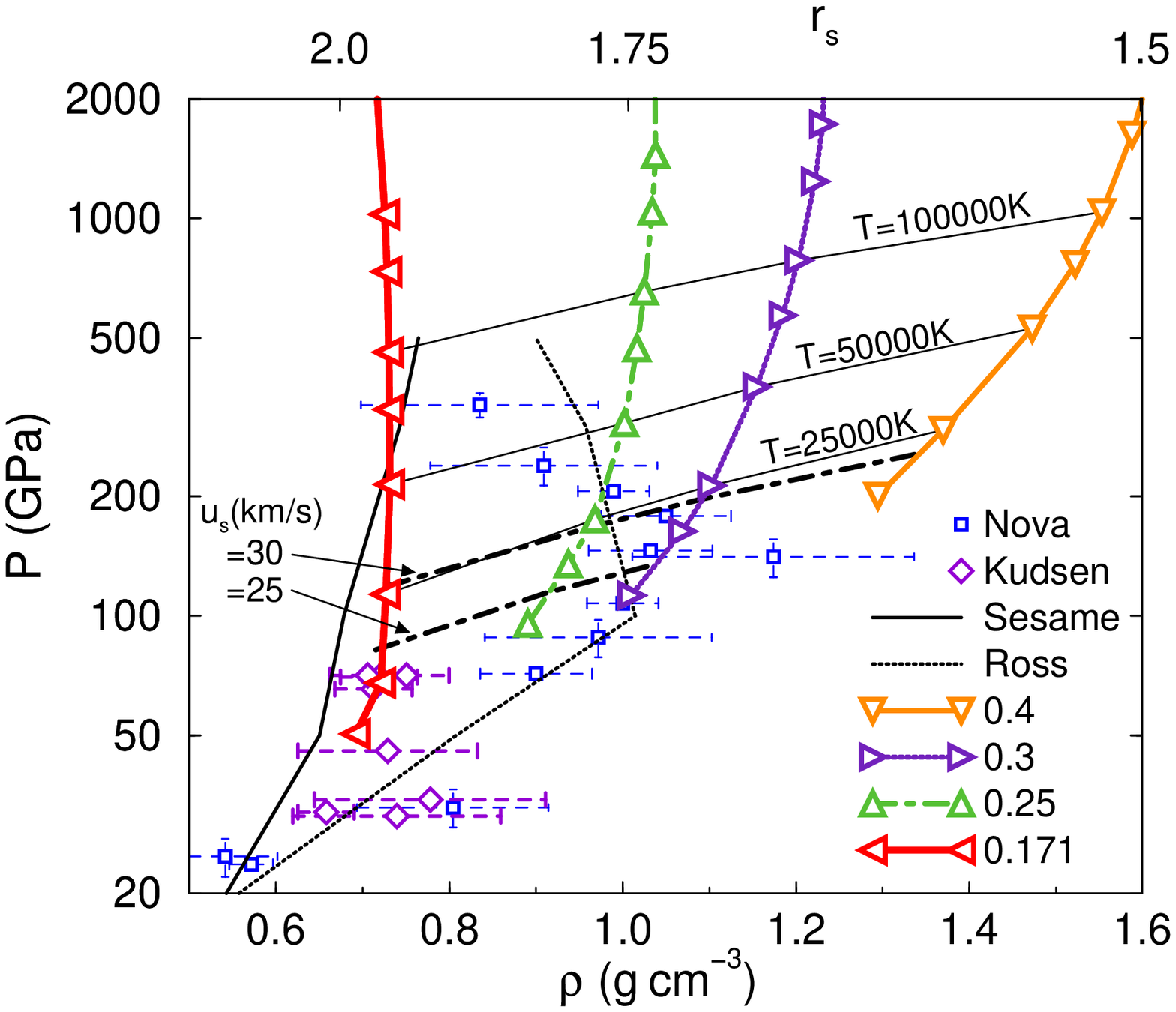}
\end{center}
\caption{\label{prho}
    	Pressure density diagram showing Hugoniot measurements
    	\cite{Si97,Co98,Kn01} and results from chemical model
    	calculations~\cite{Ke83,Ro98} starting from the uncompressed
    	liquid deuterium state $(\rho_0=0.171\,\rm g\,cm^{-3})$. The
    	triangles indicate Hugoniot curves from PIMC simulations for
    	different initial densities $\rho_0$ given in the legend in units
    	of g$\,$cm$^{-3}$. The thin solid lines are isotherms, the
    	dash-dotted lines represent final shock states generated with
    	the same shock velocity ($u_s=25$ and 30 km/s) starting from
    	different initial compressions.}
\end{figure}

The precompression in a DAC allows one to probe regions of higher
density as shown in Figs.~\ref{phase} and~\ref{prho}. The different
curves show predictions based on PIMC simulations for different
initial compressions. The densities $\rho_0=0.25$ and 0.30$\gcc$ were
chosen to approximately represent conditions in a first set of
experiments using laser driven shocks in samples precompressed in a
DAC~\cite{cellier}. The range of initial pressures and densities is
limited by the currently available laser power, which limits the size
of diamonds being used and therefore the maximum achievable
precompression.

The equation of state from our PIMC simulations has been used to
derive the different Hugoniot curves in Fig.~\ref{prho}. Even a
relatively small initial compression to $\rho_0=0.25\gcc$ yields a
maximum final density of 1$\gcc$, which is about 6 times the
uncompressed initial density of $0.171\gcc$ and also close to the
densities predicted by the Nova experiments. However, it is expected
that the maximum in compression will be at much higher pressures of
about 1500 GPa, which is one order of magnitude higher than the
maximum in the uncompressed case, which is estimated to be at
100 to 200 GPa.

Fig.~\ref{phase} shows different Hugoniot curves for initial density
up to $\rho_0=0.6\gcc$ predicting final shock densities of up to
$2.4\gcc$. The maximum compression was always close to 4-fold the
initial density. An unexpected increase in compression similar to the
predictions of the Nova experiments has not been observed for any
initial compression. In fact, the maximum compression ratio actually
decreased slightly from 4.3-fold in the uncompressed case to about
4.1-fold for $\rho_0=0.6\gcc$.

It remains to be seen which of the characterized conditions of dense
deuterium will be realized in future experiments. It depends on the
experimental limitation such as the power of the shock drive and the
pressures that can be reached in the DAC. For example, if the
sustainable shock velocity is limited to 30$\kms$, only pressures up
to about 200 GPa can be reached if $\rho_0 \leq 0.4\gcc$. The
pressure-density points obtainable at constant shock velocity are
shown in Fig.~\ref{prho}. The locus of those points is very similar to
the isotherm of $T=25\,000\,\K$.

Much higher temperatures and pressures can be reached in a shock
reverberation measurement using the same initial shock velocity.
Comparing the precompressed single shock experiment with the double
shock experiment using an aluminum back plate, one finds that an initial
compression to $\rho_0 \approx 0.37\gcc$ leads to a primary Hugoniot
curve very similar to the secondary Hugoniot curve in the double shock
experiment (compare Fig.~\ref{phase} and \cite{Mi01}). However, for a
fixed primary shock velocity, the states reached in the double shock
experiment are at about twice the temperature and 1.8 times the
pressure compared to the compressed single shock experiment with
$\rho_0 \approx 0.37\gcc$ using the same primary shock velocity. A more
detailed discussion of these curves is expected to follow when the
first experimental data become available.

In conclusion, the EOS from PIMC simulations of dense deuterium has
been used to compute shock Hugoniot curves for different initial
precompressions. The study makes precise predictions solely based
first principles calculations to compare with shock experiments using
deuterium samples, which have been precompressed in DACs. The main
purpose of this work is to provide these curves as a {\em reference}
for the interpretation of future experiments as the precompression
techniques is further advanced.

\ack

We thank G.W.~Collins for telling us about initial conditions in the
experiment and acknowledge helpful discussions with R.~Cauble and
E.L.~Pollock. This work was performed under the auspices of the
U.S. Department of Energy by University of California Lawrence
Livermore National Laboratory under contract No. W-7405-Eng-48.

\section*{References}


\begin{thebibliography}{10}

\bibitem{Si97}
I.~B. {Da~Silva}, P.~Celliers, G.~W. Collins, K.~S. Budil, N.~C. Holmes,
  W.~T.~Jr. Barbee, B.~A. Hammel, J.~D. Kilkenny, R.~J. Wallace, M.~Ross,
  R.~Cauble, A.~Ng, and G.~Chiu.
\newblock {\em Phys. Rev. Lett.}, {\bf 78}:783, 1997.

\bibitem{Kn01}
M.~D. Knudson, D.~L. Hanson, J.~E. Bailey, C.~A. Hall, J.~R. Asay, and W.~W.
  Anderson.
\newblock {\em Phys. Rev. Lett.}, 87:225501, 2001.

\bibitem{Co98}
G.~W. Collins, L.~B.~Da Silva, P.~Celliers, D.~M. Gold, M.~E. Foord, R.~J.
  Wallace, A.~Ng, S.~V. Weber, K.~S. Budil, and R.~Cauble.
\newblock {\em Science}, {\bf 281}:1178, 1998.

\bibitem{Ke83}
G.~I. Kerley.
\newblock In J.~M. Haili and G.~A. Mansoori, editors, {\em Molecular Based
  Study of Fluids}, page 107. ACS, Washington DC, 1983.

\bibitem{Ro98}
M.~Ross.
\newblock {\em Phys. Rev. B}, {\bf 58}:669, 1998.

\bibitem{Le97}
T.~J. Lenosky, J.~D. Kress, and L.~A. Collins.
\newblock {\em Phys. Rev. B}, {\bf 56}:5164, 1997.

\bibitem{MC00}
B.~Militzer and D.~M. Ceperley.
\newblock {\em Phys. Rev. Lett.}, 85:1890, 2000.

\bibitem{Mo00}
A.N. Mostovych, Y.~Chan, T.~Lehecha, A.~Schmitt, and J.D. Sethian.
\newblock {\em Phys. Rev. Lett.}, 85:3870, 2000.

\bibitem{Mi01}
B.~Militzer, D.~M. Ceperley, J.~D. Kress, J.~D. Johnson, L.~A. Collins, and
  S.~Mazevet.
\newblock {\em Phys. Rev. Lett.}, 87:275502, 2001.

\bibitem{DW02}
M.~W.~C. Dharma-wardana and F.~Perrot.
\newblock {\em Phys. Rev. B}, 66:014110, 2002.

\bibitem{GG02}
F.~Gygi and G.~Galli.
\newblock {\em Phys. Rev. B}, 65:220102, 2002.

\bibitem{Kn02}
M.~Knaup.
\newblock PhD thesis, University of Erlangen, Germany, 2002.

\bibitem{Ze66}
Y.~B. Zeldovich and Y.~P. Raizer.
\newblock {\em Physics of Shock Waves and High-Temperature Hydrodynamic
  Phenomena}.
\newblock Academic Press, New York, 1966.

\bibitem{Ce95}
D.~M. Ceperley.
\newblock {\em Rev. Mod. Phys.}, 67:279, 1995.

\bibitem{Ce96}
D.~M. Ceperley.
\newblock In Ed.~K. Binder and G.~Ciccotti, editors, {\em Monte Carlo and
  Molecular Dynamics of Condensed Matter Systems}. Editrice Compositori,
  Bologna, Italy, 1996.

\bibitem{Ce91}
D.~M. Ceperley.
\newblock {\em J. Stat. Phys.}, 63:1237, 1991.

\bibitem{BM00}
B.~Militzer.
\newblock PhD thesis, University of Illinois at Urbana-Champaign, 2000.

\bibitem{PC94}
C.~Pierleoni, D.M. Ceperley, B.~Bernu, and W.R. Magro.
\newblock {\em Phys. Rev. Lett.}, {\bf 73}:2145, 1994.

\bibitem{Ma96}
W.~R. Magro, D.~M. Ceperley, C.~Pierleoni, and B.~Bernu.
\newblock {\em Phys. Rev. Lett.}, {\bf 76}:1240, 1996.

\bibitem{MC01}
B.~Militzer and D.~M. Ceperley.
\newblock {\em Phys. Rev. E}, 63:066404, 2001.

\bibitem{cellier}
G. W. Collins and P. Cellier used the Omega laser to launch a shock into
  precompressed deuterium samples at room temperature and kilobar pressures.
  Initial densities were approximately 0.25 and 0.30$\gcc$. Private
  communication, 2002.

\end{thebibliography}

\end{document}